# Robust magnetic order of Ce 4f-electrons coexisting with superconductivity in $CeFeAsO_{1-x}F_x$

T. Shang[1], L. Jiao[1], J. Dai[1] and H. Q. Yuan[1,*]

[1]*Center for Correlated Matter and Department of Physics, Zhejiang University, Hangzhou, Zhejiang 310027, China*

F. F. Balakirev[2]

[2]*NHMFL, Los Alamos National Laboratory, MS E536, Los Alamos, NM 87545, USA*

W. Z. Hu[3], N. L. Wang[3]

[3]*Beijing National Laboratory for Condensed Matter Physics, Institute of Physics, Chinese Academy of Science, Beijing 10080, China*

The electrical resistance of $CeFeAsO_{1-x}F_x$ (x = 0.06 and 0.08) has been measured in a magnetic field up to 40T. At zero field, the sample with x = 0.06 shows a structural phase transition around $T_S \approx$ 100K, followed by a spin-density-wave (SDW) transition around $T_{SDW} \approx$ 30K. For x = 0.08, the structural phase transition is suppressed down to $T_S \approx$ 60K without a clear anomaly associated with the Fe-SDW transition, and superconductivity shows up at $T_C \approx$ 25K. At lower temperatures, both samples show a clear resistive peak around $T_N \approx$ 4K, which is associated with the antiferromagnetic (AFM) transition of Ce-4f electrons. Strikingly, zero resistance is recovered upon further lowering temperature below $T_N$ for x = 0.08. Moreover, we found that the AFM transition of Ce 4f-electrons at 4K hardly changes with applying a magnetic field up to 40T, even in the case of x = 0.08, where superconductivity has been partially suppressed at such a large field.

# I. INTRODUCTION

The recent discovery of superconductivity in the F-doped LaFeAsO compounds has led to tremendous efforts on searching for new materials with higher superconducting transition temperature and revealing the superconducting mechanism in these materials [1]. Substitution of La with other rare-earth elements e.g. Ce, Sm, Nd, Pr or Gd in the *Ln*Fe*Pn*O family (*Ln*: Lanthanoids, *Pn*: As or P) has raised the superconducting transition temperature up to 56 K [2-7]. In these materials, superconductivity emerges in close proximity to the SDW instability [8,9], resembling those of the high $T_C$ cuprates or the heavy fermion compounds, which provides an alternative example for studying the interplay of superconductivity and magnetism. Furthermore, the *Ln*-based iron pnictides also exhibit some striking features associated with the 4f-electrons. For instance, CeFeAsO undergoes an almost degenerate SDW transition of Fe 3d-electrons and structural phase transition around 150 K, followed by an AFM transition of Ce 4f-electrons at $T_N$ ≈4 K[3,8]. In CeFe$_{1-x}$Co$_x$AsO system [10], it is found that superconductivity sets in at 0.05< x <0.2, showing a maximum $T_C$ ≈ 14 K. At lower temperatures, the AFM order of Ce 4f-electrons is rather robust against the Co-substitution. Experiments of neutron scattering and muon spin relaxation have shown evidence of a moderate 3d-4f coupling in the CeFeAsO[11]. How the Ce 4f-electrons interact with the conducting FeAs layers and whether the 4f-electrons have a strong influence on superconductivity in the F-doped CeFeAsO remain a puzzle and demand further investigations.

In this paper, we report measurements of the electrical resistance R(*T*, *B*) of CeFeAsO$_{1-x}$F$_x$ (x = 0.06 and 0.08) in a magnetic field up to 40T. It is found that the magnetic order of Ce 4f-electrons is highly robust against magnetic field, even in the superconducting state. Our results indicate that the Ce-4f electrons are weakly coupled to the 3d-electrons physics in CeFeAsO$_{1-x}$F$_x$. The magnetic transition of Ce-4f electrons gives a perturbation to superconductivity, leading to the reentrance of superconductivity below $T_N$.

## II. EXPERIMENTS AND DISCUSSION

Polycrystalline samples of $CeFeAsO_{1-x}F_x$ are synthesized by a solid state reaction method [3,10]. The resulting samples were characterized by powder x-ray diffraction (XRD) at room temperature, which identifies a single phase with a tetragonal ZrCuSiAs-type structure (space group P4/nmm). The electrical resistance was measured with a typical four-point method using the facilities of pulsed magnetic fields at Los Alamos. Temperature dependence of the resistance at zero field was measured with a Lakeshore resistance bridge.

Fig. 1 plots the electrical resistance $R(T)$ of $CeFeAsO_{1-x}F_x$ (x = 0.06 and 0.08) at zero-field, inset of which shows the phase diagram as a function of F concentration, determined from the neutron scattering measurements [8]. One can see that Ce becomes antiferromagnetically ordered below $T_N \approx 4K$ in the low doping region. The samples we have measured (x = 0.06, 0.08) are marked by the arrows in the phase diagram. For x = 0.06, the resistance $R(T)$ dramatically changes slope around $T_S \approx 100K$ and $T_{SDW} \approx 30K$, which are ascribed to the structural phase transition and the Fe-SDW transition, respectively. While for x = 0.08, the structural phase transition is suppressed to $T_S \approx 60K$ and superconductivity emerges at $T_C \approx 25K$. At low temperature around $T_N \approx 4K$, a clear resistive peak associated with the AFM order of Ce 4f-electrons, can be seen in both x=0.06 and 0.08. Remarkably, the resistance shows a superconducting reentrance below $T_N$ for x = 0.08. Similar reentrant behavior was previously reported in $LnNi_2B_2C$ (Ln = Ho, Er, Tm) compounds, which is associated with the AFM order of Ln ions[12].

Temperature dependence of the electrical resistance at several magnetic fields, obtained from the $R(B)$ data at various temperatures, is plotted in Fig.2 for $CeFeAsO_{1-x}F_x$. As a reference, we also include the resistance $R(T)$ measured while cooling down the samples at zero field (see the black curves). For both x = 0.06 and x = 0.08, the AFM order of Ce 4f-electrons is extremely robust against the external magnetic fields, which transition temperature hardly shifts up to a field of B = 40T. Moreover, the sample with x = 0.06 shows significant magnetoresistance at low temperatures, being consistent with our measurements on the parent compound [13]. For the superconducting sample CeFeAsO$_{0.92}$F$_{0.08}$ ($T_C \approx$ 25K, $B$ = 0T), superconductivity is clearly suppressed to lower temperatures when applying a magnetic field ($T_C \approx$ 15K, $B$ = 40T), but the resistive peak associated with Ce-AFM in the superconducting state stays unchanged. It is surprising that a magnetic field of 40T has little effect on the low-lying magnetic order of Ce-4f electrons. This suggests that the characteristic energy scale of the Ce magnetism, which is mainly determined from the frustrated superexchange interactions via Ce-O-Ce and Ce-As-Ce paths [14], is strong in spite of the relatively low ordering temperature. The 3d-electrons pairing instability does not correlate to the magnetism of Ce 4f-electrons, but the pair-breaking associated with the magnetic order is strong enough to bring a resistive reentrant behavior. One should point out that measurements of CeFeAsO single crystal demonstrated strong easy-plane anisotropy [15]. The AFM order of Ce is more rapidly suppressed by an in-plane magnetic field, but is similarly robust against magnetic field for B//c as we observed in the F-doped compounds here. It would be desired to study the magnetic anisotropy of Ce in the charge-carrier doped CeFeAsO, which may shed insight on the interactions of 4f and 3d electrons, and therefore the interplay of magnetism and superconductivity in iron pnictides. For this purpose, high quality single crystals of CeFeAsO$_{1-x}$F$_x$ and other sister compounds are demanded.

## III. CONCLUSION

To summarize, we have studied the effect of magnetic field on the superconductivity and Ce-AFM order in CeFeAsO$_{1-x}$F$_x$. It is found that the magnetic order of Ce 4f-electrons is highly robust against magnetic field, indicating that the magnetic interactions of Ce 4f-electrons have a large characteristic energy scale in spite of its low-lying ordering temperature. The coupling between the 3d- and 4f-electrons seems to be weak in these compounds. The magnetic transition of Ce-4f electrons does cause a resistive peak in the superconducting state of CeFeAsO$_{1-x}$F$_x$ (x=0.08), but superconductivity reenters below the Néel transition ($T_N \approx$ 4K) of Ce-4f electrons.

## ACKNOWLEDGEMENT

This work was supported by NSFC, the National Basic Research Program of China (973 program), the PCSIRT of the Ministry of Education of China, Zhejiang Provincial Natural Science Foundation of China and the Fundamental Research Funds for the Central Universities. Work at NHMFL-LANL is performed under the auspices of the National Science Foundation, Department of Energy and State of Florida.
## REREFENCES

[1] Y.Kamihara et al., J. Am. Chem. Soc. **130**, 3296 (2008).
[2] X.H.Chen et al., Nature **453**, 761 (2008).
[3] G. F. Chen et al., Phys. Rev. Lett. **100**, 247002 (2008).
[4] Z. A. Ren et al., Chin. Phys. Lett. **25**, 2215 (2008).
[5] C. Wang et al., Europhys. Lett. **83**, 67006 (2008).
[6] C. R. Rotundu et al., Phys. Rev. B .**80**, 144517 (2009).
[7] H. Kito et al., J. Phys. Soc. Jpn. **77**, 3707 (2008).
[8] J. Zhao et al., Nature Materials **7**, 953 (2008).
[9] H.Okada et al., J. Phys. Soc. Jpn. **77**, 3712 (2008).
[10] T.Shang et al., unpublished.
[11] H. Maeter et al., Phys. Rev. B. **80**, 094524 (2009).
[12] H. Eisaki et al., Phys. Rev. B. **50**, 647 (1994).
[13] H. Q. Yuan et al., Journal of Physics: Conference Series.**273**, 012110 (2011).
[14] J. Dai et al., Phys. Rev. B.**80**, 020505(R) (2009).
[15] A.Jesche et al., New J. Phys. **11**, 103050 (2009).

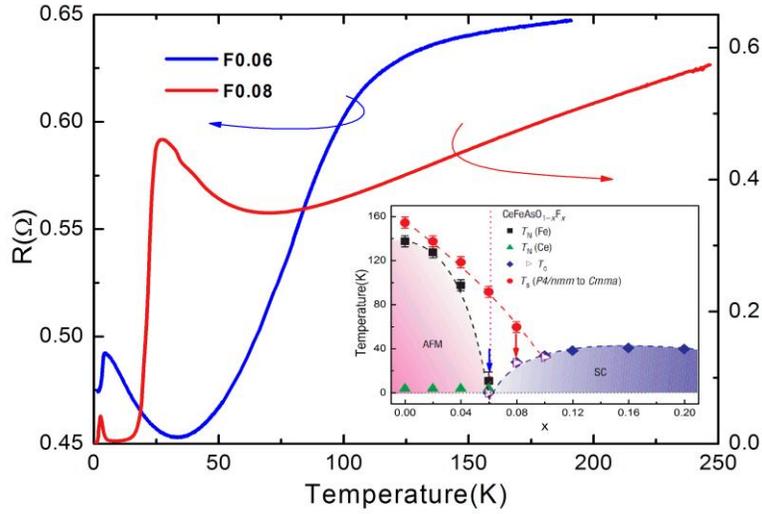

**Fig.1.** Temperature dependence of the electrical resistance at zero field for CeFeAsO$_{0.94}$F$_{0.06}$ (blue) and CeFeAsO$_{0.92}$F$_{0.08}$ (red). The inset shows the phase diagram of CeFeAsO$_{1-x}$F$_x$ as a function of F concentration (x) obtained in reference [8]. The arrows in the inset denote the two samples studied here.

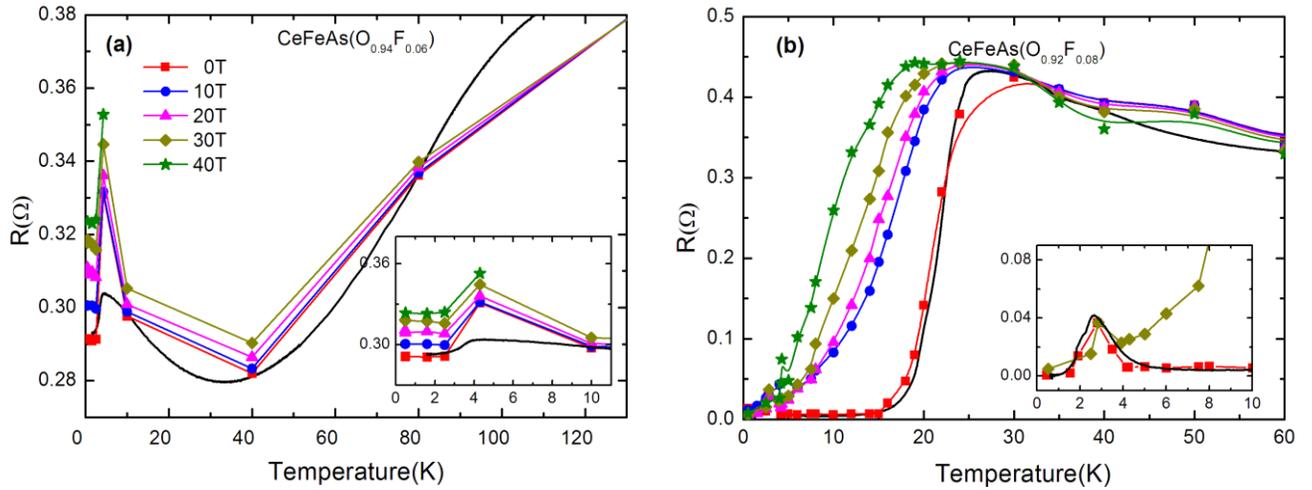

**Fig.2.** Temperature dependence of the resistance at various magnetic fields up to 40 T for CeFeAsO$_{1-x}$F$_x$ (x = 0.06 and 0.08). The insets of (a) and (b) plot the data below 10 K. The black curves show the zero field resistance measured while cooling down the samples from room temperature, and the symbols represent the data measured in pulsed fields.